\documentclass[11pt]{article}
\newsavebox{\PSLASH}
\sbox{\PSLASH}{$p$\hspace{-1.8mm}/}

\input{amssym}
\def\be{\begin{equation}}
\def\ee{\end{equation}}
\def\ba{\begin{eqnarray}}
\def\ea{\end{eqnarray}}
\begin{document}
\title{\large \bf A Note on the Generalized Friedmann Equations for a Thick Brane}
\author{ Samad Khakshournia$^{1}$
\footnote{Email Address: skhakshour@aeoi.org.ir}\\
$^{1}$ Nuclear Science and Technology Research Institute
(NSTRI),Tehran, Iran }\maketitle
\[\]
 \[\]
 \[ \]
\begin{abstract}
Within our thick brane approach previously used to obtain the
cosmological evolution equations on a thick brane embedded in a
five-dimensional Schwarzschild Anti-de Sitter spacetime it is
explicitly shown that the consistency of these equations with the
energy conservation equation requires that, in general, the
thickness of the brane evolves in time. This varying brane
thickness entails the possibility that both Newton's gravitational
constant $G$ and the effective cosmological constant $\Lambda_4$
are time dependent.

\end{abstract}

\newpage

\section{\large \bf Introduction}
Cosmological models in which the observed universe is realized as
a four-dimensional thin brane embedded in a higher dimensional
spacetime (bulk) are of current interest. In these theories the
cosmological evolution on the brane is described by the effective
Friedmann equations that incorporates nontrivially the effects of
the bulk. The most important feature that distinguishes brane
cosmology from the standard scenario is the fact that at high
energies Friedmann equation modifies by an extra term quadratic in
the energy density of matter on the brane \cite{Bin99}. Several
features and variations of this scenario including thick brane
configurations, in particular, in the cosmological context have
been considered. In the approach used in Ref. \cite{Moun} to
derive generalized Friedmann equations, the four-dimensional
effective brane quantities are obtained by integrating the
corresponding five-dimensional ones along the extra-dimension over
the brane thickness. These cosmological equations describing a
brane of finite thickness interpolate between the case of an
infinitely thick brane corresponding to the familiar Kaluza-Klein
picture and the opposite limit of an infinitely thin brane giving
the unconventional Friedmann equation, where the energy density
enters quadratically. The latter case is then made compatible with
the conventional cosmology at late times by introducing and fine
tuning a negative cosmological constant in the bulk and an
intrinsic positive tension in the brane \cite{whole1}. Authors in
Ref. \cite{Nav05} considered a thick codimension 1 brane including
a matter pressure component along the extra dimension in the
energy-momentum tensor of the brane. By integrating the 5D
Einstein equations along the fifth dimension, while neglecting the
parallel derivatives of the metric in comparison with the
transverse ones, they wrote the equations relating the values of
the first derivatives of the metric at the brane boundary with the
integrated componentes of the brane energy-momentum tensor. These,
so called matching conditions were then used to obtain the
cosmological evolution of the brane which is of a non-standard
type, leading to an accelerating universe for special values of
the model parameters. On the other hand the cosmological
implications of brane world scenario in which fundamental physical
constants are variable functions of time have been investigated
\cite{Gvar1,Gvar2}. For instance, in Ref. \cite{Gvar1} was shown
that the introduction of a scalar function varying with time in
the brane-world theory yields a number of cosmological models
which do not admit constant values for the Newton's gravitational
coupling $G$ and the effective cosmological term $\Lambda_4$.\\
We are interested in the effects arising from taking into account
the thickness of the brane. In this way, we have presented a
general equation to study the dynamics of codimension one brane of
finite thickness immersed in an arbitrary bulk spacetime. That was
obtained in a general setting by imposing the Darmois junction
conditions on the brane boundaries with the two embedding
spacetimes and then using an expansion scheme for the extrinsic
curvature tensor at the brane boundaries in terms of the proper
thickness of the brane. Using this approach we have derived the
generalized Friedmann equations written up to the first order of
the brane proper thickness governing the cosmological evolution of
a thick brane embedded in a five-dimensional
Schwarzschild Anti-de Sitter spacetime with a $Z_{2}$ symmetry
\cite {ghkhm}.\\
The aim of this note is to address the presence of the brane with
a time dependent thickness as a general feature of our thick brane
cosmological solutions. \\
In section 2 we give a brief review of the generalized
cosmological equations as obtained in Ref. \cite{ghkhm}. In
section 3 we derive an evolution equation for the brane thickness
as required from the self-consistency of our equations. In section
4 conclusion is presented.
\section{\large \bf  Generalized Cosmological Equations}
We consider a thick brane embedded in a five-dimensional
Schwarzschild Anti-de Sitter spacetime with the metric
\begin{equation}\label{metout}
ds^2=-f(r)dT^2+\frac{dr^2}{f(r)}+r^2d\Omega_{k}^{2},
\end{equation}
with
\begin{equation}\label{f}
f(r)=k-\frac{\Lambda}{6}r^2-\frac{C}{r^2},
\end{equation}
where $d\Omega_{k}^{2}$ is the metric of the 3D hypersurfaces
$\Sigma$ of constant curvature that is parameterized by $k=0, \pm
1$; $\Lambda$ is the bulk cosmological constant, and constant $C$
is identified with the mass of a black hole located at $r=0$.  We
then take the following ansatz for the metric of the brane being
written in a Gaussian normal coordinate system in the vicinity of
the core of the thick brane situated at $y=0$
\begin{equation}\label{metin}
ds^2=-n^2(t,y)dt^2+dy^2+a^2(t,y)d\Omega_{k}^{2},
\end{equation}
where $n^2(t,y)$ and $a^2(t,y)$ are some unknown functions, and
$y$ is the normal coordinate of the extra dimension. Compatibility
with the cosmological symmetries requires that the energy-momentum
tensor of the matter content in the brane takes the simple form
\begin{equation}\label{stress}
T^{\mu}_{\nu}=(-\rho,P_L,P_L,P_L,P_T),
\end{equation}
where the energy density $\rho$, the longitudinal pressure $P_L$,
and the transverse pressure $P_{T}$  are functions of $t$ and $y$.\\
The generalized first Friedmann equation written up to first order
of the brane thickness takes the following form within our
formalism \cite {ghkhm}
\begin{equation}\label{f1}
\sqrt{f_{0}+\dot{a}_{0}^2}=\frac{w}{a_{0}}\left(\frac{-\Lambda}{3}a_{0}^2
+\frac{\kappa^2}{3}\rho_{0}a_{0}^2\right),
\end{equation}
where the subscript "0" means evaluation at the center of the
thick brane and  the dot stands for the derivative with respect to
the proper time $\tau$ on the brane center. Taking the square of
Eq. (\ref{f1}), substituting the expression (\ref{f}) and
rearranging, we arrive at the following equation
\begin{equation}\label{eqs}
H_{0}^2+\frac{k}{a_{0}^2} = \frac{8\pi
G}{3}\varrho+\frac{\kappa^4}{36}\varrho^2
+\frac{\Lambda_4}{3}+\frac{C}{a_{0}^4},
\end{equation}
where $H_0=\frac{\dot{a}_{0}}{a_{0}}$, and the effective
four-dimensional energy density $\varrho$ associated to the
five-dimensional energy density $\rho$ has been defined as
\begin{equation}\label{rho}
\varrho=\int_{-w}^{w} \rho dy\simeq 2w\rho_{0}+O(w^2),
\end{equation}
and the following identifications have been made
\begin{eqnarray}\label{effect}
\frac{\Lambda_4}{3}=\frac{\Lambda}{6}
+\frac{w^2\Lambda^2}{9},\hspace{1cm} 8\pi G =
\frac{\kappa^2w(-\Lambda)}{3}.
\end{eqnarray}
We see that there is a linear in addition to a quadratic term in
the matter density, due to the non-vanishing of the thickness $w$.
There is no need of introducing an ad hoc tension for the brane,
and splitting it from the matter density on the brane.\\
The generalized second Friedmann equation written up to first
order of the brane thickness takes the following form within our
formalism \cite {ghkhm}
\begin{eqnarray}\label{acc}
\frac{\frac{\ddot{a}_{0}}{a_{0}}-\frac{\Lambda}{6}
+\frac{C}{a_{0}^4}}{\sqrt{f_{0}+
\dot{a}_{0}^2}}&=&\frac{-w}{a_{0}}\left(
\frac{2\kappa^2}{3}\left(\rho_{0}
+\frac{3}{2}P^{0}_{L}-P^{0}_{T}+\frac{3}{a_{0}^4}\tilde{P_T}\right)
+\frac{\Lambda}{6}+\frac{3E}{a_{0}^4} +\frac{3C}{a_{0}^4}\right),
\end{eqnarray}
where $E>0$ is an integration constant, $P^{0}_{L}=P_{L}(t,y=0)$,
and $P^{0}_{T}=P_{T}(t,y=0)$, and we have also defined
$\tilde{P_T}\equiv\int_{0}^{\tau} P^{0}_T\:a^4_0\:H_0d\tau$. Note
now that the time component of the covariant derivative of the
brane energy-momentum tensor (\ref{stress}), using the metric
(\ref{metin}), leads to the familiar energy conservation condition
on the core of the thick brane:
\begin{equation}\label{conser}
\dot{\rho_{0}}+3H_{0}(\rho_{0}+P^0_{L})=0.
\end{equation}
Defining the four-dimensional effective quantities associated to
the five-dimensional longitudinal and transverse pressures $P_{L}$
and $P_{T}$ in the form
\begin{equation}\label{pl}
 p_{L}=\int_{-w}^{+w}P_{L}\:dy\simeq 2wP^0_{L}+O(w^2),
 \end{equation}
\begin{equation}\label{pt}
 p_{T}=\int_{-w}^{+w}P_{T}\:dy\simeq 2wP^0_{T}+O(w^2),
 \end{equation}
Let us assume the arbitrary effective equations of state of the
form
\begin{equation}\label{sta}
 p_{L}=\omega_{L}\varrho,\hspace{1cm}p_{T}=\omega_{T}\varrho,
\end{equation}
with constants $\omega_{L}$ and  $\omega_{T}$. The conservation
equation (\ref{conser}) can then be integrated with the result as
usual
\begin{equation}\label{conser2}
\rho_{0}=\rho_{i}a_{0}^{-3(1+\omega_{L})},
\end{equation}
where $\rho_{i}$ is a constant. Then $\tilde{P_T}$ can be computed
as
\begin{equation}\label{tildePevol}
\tilde{P_T}=\frac{\rho_{i}\omega_{T}}{1-3\omega_{L}}a_{0}^{1-3\omega_{L}}.
\end{equation}
\section{\large \bf Evolution of Brane Thickness}
Let us now write down the time derivative of the first Fridmann
equation (\ref{f1}) while allowing the brane thickness $w$ to
evolve in time. We get
\begin{eqnarray}\label{deriv}
\frac{\frac{\ddot{a}_{0}}{a_{0}}-\frac{\Lambda}{6}
+\frac{C}{a_{0}^4}}{\sqrt{f_{0}+
\dot{a}_{0}^2}}&=&\frac{-w}{a_{0}}\left(
\frac{2\kappa^2}{3}\left(\rho_{0} +\frac{3}{2}P^{0}_{L}\right)
+\frac{\Lambda}{3}\right)+\frac{\dot{w}}{\dot{a}_{0}}
\left(-\frac{\Lambda}{3}+\frac{\kappa^2}{3}\rho_{0}\right),
\end{eqnarray}
where the energy conservation equation (\ref{conser}) has been
used. Comparing this with the second Fridmann equation (\ref{acc})
we infer
\begin{eqnarray}\label{deriv3}
\frac{\dot{w}}{w}=\frac{-H_{0}}{\frac{-\Lambda
}{3}+\frac{\kappa^2}{3}\rho_{0}}\left(\frac{2\kappa^2\rho_{0}}{3}
\left(\frac{3\omega_{T}}{1-3\omega_{L}}-\omega_{T}\right)+\left(\frac{3E}{a_{0}^4}
+\frac{3C}{a_{0}^4}+\frac{-\Lambda}{6}\right)\right),
\end{eqnarray}
where we have used Eqs. (\ref{rho}), (\ref{pl}), (\ref{pt}),
(\ref{sta}), (\ref{conser2}), and (\ref{tildePevol}). This is an
evolution equation for the brane thickness indicating, in general,
the presence of the brane with a time dependent thickness.
Depending on the magnitude and sign of the different terms within
the bracket in Eq. (\ref{deriv3}), the brane thickness may
increase or decrease with time. For instance, we observe that for
$\omega_{T}$=0, the brane thickness extended along the extra
dimension always decreases with time while its ordinary spacial
dimensions expand if $k=0$, or $-1$. On the other hand for
$\omega_{T}<0$ leading to an accelerating brane cosmology at late
time \cite{ghkhm}, from Eq. (\ref{deriv3}) one can see that it is
possible to have an increasing brane thickness with time.
Furthermore, as the equation (\ref{deriv3}) poses, the parameters
within the bracket must be finely tuned to yield a time constant
brane thickness.
\section{\large \bf Conclusion}
In this note we showed that the requirement of compatibility our
thick brane cosmological equations written up to the first order
of the brane thickness with the familiar energy conservation
equation leads to an evolution equation for the brane thickness.
In the absence of the pressure along the extra dimension in the
brane energy-momentum tensor the thickness of the brane decreases
with time while at late time a negative transverse pressure can
lead to an increase in the brane thickness. We have realized that
our two generalized Friedmann equations (\ref{f1}) and (\ref{acc})
together with the energy conservation equation (\ref{conser}) form
the three independent equations to determine the evolution of the
three unknown functions $a_{0}$, $\rho_{0}$, and $w$ once one gets
the equations of state (\ref{sta}) for brane matter. According to
the identifications made in (\ref{effect}) a time dependent brane
thickness induces time variations in the Newton's gravitational
coupling $G$ and the effective cosmological term $\Lambda_4$. This
suggests our thick brane cosmological model as a general framework
in which one enables to study the simultaneous variation of $G$
and $\Lambda_4$.
\section{Acknowledgments}
The author would like to thank Professor R. Mansouri and S.
Ghassemi for fruitful comments that led to the development of this
work.

\end{document}